\documentclass[aps,prx,twocolumn,amsmath,amssymb,superscriptaddress,floatfix]{revtex4-1}

\usepackage{graphicx}
\usepackage{multirow}
\usepackage{color}
\usepackage{bm}
\usepackage{hyperref}

\renewcommand{\Vec}[1]{{\bm{#1}}}
\def\infinity{\infty}
\def\t#1{\textrm{#1}}

\def\braket#1{\langle #1 \rangle}
\def\n{\nonumber \\ }

\begin{document}

\title{
Scaling laws for nonlinear electromagnetic responses of Dirac fermions
}

\author{Takahiro Morimoto}
\affiliation{Department of Physics,
University of California, Berkeley, CA 94720}
\affiliation{RIKEN Center for Emergent Matter Science 
(CEMS), Wako, Saitama, 351-0198, Japan}
\author{Naoto Nagaosa}
\affiliation{RIKEN Center for Emergent Matter Science 
(CEMS), Wako, Saitama, 351-0198, Japan}
\affiliation{Department of Applied Physics, The University of 
Tokyo, Tokyo, 113-8656, Japan}

\date{\today}

\begin{abstract}
We theoretically propose that the Dirac fermion in two-dimensions 
shows the giant nonlinear responses to electromagnetic fields in terahertz region.
A scaling form is obtained for the current and magnetization as functions of
the normalized electromagnetic fields $E/E_\omega$ and $B/B_\omega$,
where the characteristic electric (magnetic) field $E_\omega$ 
($B_\omega$) depends on the frequency $\omega$ as
$\hbar\omega^2/ev_F$ ($\hbar\omega^2/ev_F^2$), and is typically of the order of  
80 V/cm ( 8 mT) in the terahertz region.
Applications of the present theory to graphene and surface state of a topological
insulator are discussed.  
\end{abstract}

% insert suggested PACS numbers in braces on next line
%72.10.-d 	Theory of electronic transport; scattering mechanisms
%73.20.-r 	Electron states at surfaces and interfaces
%78.67.-n 	Optical properties of low-dimensional, mesoscopic, and nanoscale materials and structures
%42.65.-k 	Nonlinear optics
\maketitle

\section{Introduction}
%\textit{Introduction ---}
Nonlinear electromagnetic responses (NLEMRs)%
~\cite{Boyd,Kishida00,Kishida01,McIver12,Junck13,Ogawa13,Ogawa14,Young12,Grinberg13}
are essential for creating efficient electric/magnetic devices 
because their functions are closely related to the nonlinearity.
In particular, the NLEMRs in the low energy region, e.g., terahertz frequency region,
is the focus of keen attention due to the development of the intensive terahertz
excitations, and the strong coupling between the free carriers 
and electromagnetic field (EMF) in 
semiconductors~\cite{Ganichev,Hoffmann,Turchinovich12,Cornet14}.
To achieve the strong nonlinearity in low energy, two conditions must be satisfied.
One is that there are enough excited states in that region, and the other is that
the strong EMF can exist in the material.
Nonlinear responses are enhanced when there exist resonant states corresponding to the frequency of the external fields
and the electric field within the material can be large by avoiding the screening effect.
If we focus on the low frequency region which is our interest in this paper,
metals could be a good candidate to show the NLEMRs for the first point,
because there are large density of states in the low
energy region. 
However, the electromagnetic field is screened by the 
conducting electrons and hence cannot be strong enough to
show the nonlinearity. 
In insulators, on the other hand, 
the energy gap $E_g$ leads to the stability of the system, 
while the energy denominator $\hbar\omega - E_g$ appearing
in the perturbation theory reduces the higher order effects, and hence strong nonlinear effects. 
The resonance at $\omega \sim E_g$ is one way to
enhance the NLEMRs, but it does not work for the
low energy responses in the THz regime.
%~\cite{Ganichev,Hoffmann,Turchinovich12,Cornet14}.
Another source of the nonlinearity is 
the symmetry breaking such as the ferroelectricity, which 
produces the giant responses by the small external field. 
For example, photovoltaic currents have recently been studied for ferroelectric materials~\cite{Young12,Grinberg13}.
However, the switching speed of ferroelectricity is usually very slow,
which deteriorates its applicability for realistic devices.
   
\begin{figure}
\begin{center}
\includegraphics[width=0.5\linewidth]{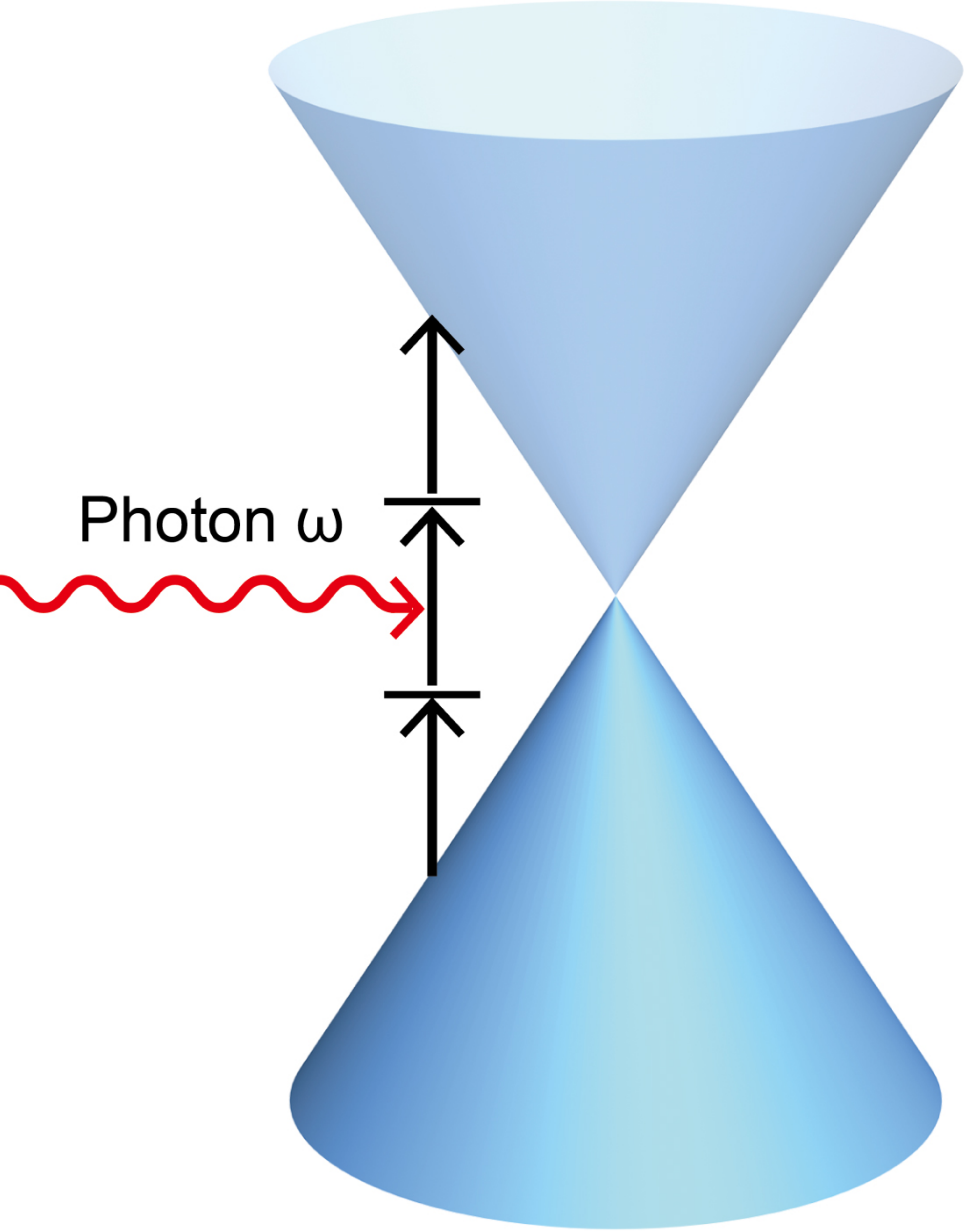}
\caption{\label{fig: WF}
Schematic picture of nonlinear responses of a Dirac fermion.
}
\end{center}
\end{figure}

From this viewpoint, the Dirac fermion having linear dispersions 
(Fig.~\ref{fig: WF})
is located between the metals 
and insulators, and has an advantage of having low energy excitations
and poor screening. Therefore, it deserves to explore 
theoretically the possible giant NLEMRs of two-dimensional Dirac fermions which are realized in condensed matters at the surface of three-dimensional topological insulators (TIs)~\cite{TI1,TI2},
and also in graphene~\cite{graphene1,graphene2}.
The screening of external fields are weak for Dirac fermions 
since the density of states vanish at zero energy.
At the same time, the linear dispersion of Dirac fermions enables that a light of any frequency matches the resonance condition,
especially in the terahertz region. 
Actually, it has been proposed that graphene is a promising candidate
for the NLEMRs~\cite{Glazov2014,Wright09,Hendry10,Mikhailov07,Mikhailov14,Paul13,Mics15}.
In addition, Dirac fermions in TIs show a nontrivial topological nature such as magneto-electric responses~\cite{TME,Essin09},
which makes TIs the attractive system for the
NLEMRs.
Since Dirac fermions in TIs are subject to the spin-orbit coupling,
NLEMRs in TIs indicate nonlinear responses of spins.
In this regard, nonlinear responses of TIs will lead to novel spintronics functions of TIs.

In this paper, we study nonlinear responses of a two-dimensional Dirac fermion and derive a scaling form for them.
This provides a unified description of various nonlinear phenomena supported by two-dimensional Dirac fermions which are realized in graphene and at the surface of TIs.

\section{Two regimes of nonlinear responses}
%\textit{Two regimes of nonlinear responses ---}
We consider the Dirac fermion at the surface of a topological insulator 
or in a graphene (neglecting the valley and spin degrees of freedom)
which is described by the Dirac Hamiltonian, 
\begin{align}
H=\hbar v_F(k_x \sigma_y - k_y \sigma_x) ,
\end{align}
where $h= 2 \pi \hbar $ is the Planck constant,
$\sigma$ are Pauli matrices acting on spin degrees of freedom,
and $v_F$ is the Fermi velocity of the Dirac fermion.
The Green's function for the Dirac fermion is given by
\begin{align}
G(i\omega)&=(i\hbar\omega-H)^{-1}=
\hbar^{-1}
\frac{i\omega+v_F(k_x \sigma_y - k_y \sigma_x)}{(i\omega)^2-v_F^2 k^2},
\end{align}
and the current operator is given by
\begin{align}
j_x=ev_F \sigma_y.
\end{align}
%Hereafter, we set $v_F=1$ because
%we can always restore $v_F$ by a dimension analysis.

\begin{figure}
\begin{center}
\includegraphics[width=0.85\linewidth]{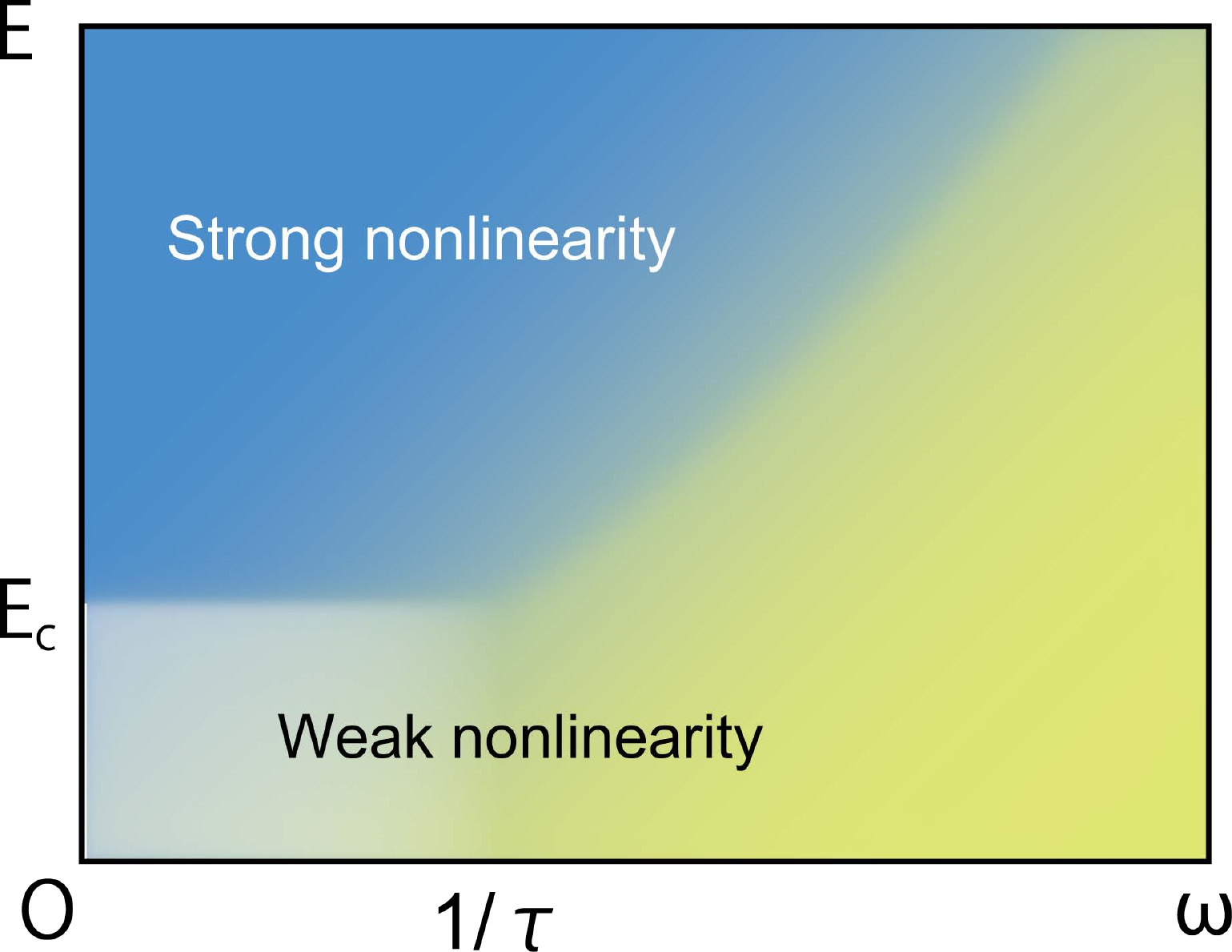}
\caption{\label{fig: phase diagram}
Two regimes of nonlinear responses.
Strong nonlinearity appears when $E>E_\omega$ and  $E>E_c$.
Weak nonlinearity appears otherwise.
Similar phase diagram applies for $\omega-B$ plane.
}
\end{center}
\end{figure}

We study NLEMRs of the Dirac fermion focusing on responses of 
uniform current $J$ and magnetization $M$  
of the Dirac fermion to external ac EMFs  $E$ and $B$. 
The NLEMRs of $J$ and $M$ obey the following scaling forms
\begin{align}
J&= \frac{e\omega^2}{v_F} f \left(
\frac{E}{E_\omega},
\frac{B}{B_\omega}
\right),
\\
M&= e\omega g \left(
\frac{E}{E_\omega},
\frac{B}{B_\omega}
\right),
\label{eq: scaling form}
\end{align}
because the Dirac fermion is gapless and the only energy scale is introduced by the frequency $\omega$ of the external EMFs.
Here, $E_\omega$ and $B_\omega$ are the typical scales of EMFs determined by the frequency of EMFs,
and $f$ and $g$ are scaling functions. 
We note that the spatially uniform limit $q \to 0$ is taken in this scaling argument.

Now there are two typical regimes of nonlinear responses depending on the strength of the EMFs and the frequency: 
(i) strong nonlinearity where $E$ ($B$) is larger than $E_\omega$ and $E_c$ ($B_\omega$ and $B_c$), and 
(ii) weak nonlinearity otherwise,
as summarized in Fig.~\ref{fig: phase diagram}.
The typical order of electric fields $E_\omega$ is determined 
by 
\begin{align}
eE_\omega v_F t=\frac{h}{t},
\end{align}
which equates the energy that a Dirac fermion driven by $E$ gains by the time $t$ and the energy scale corresponding to $t$.
The typical order of magnetic fields $B_\omega$ is determined by
\begin{align}
(v_F t)^2 B_\omega=\frac{h}{2e},
\end{align}
which equates the magnetic flux in an area $(v_F t)^2$ to a flux quantum.
These two equations indicate that the nonlinear effects 
are qualitatively different when the EMFs are larger or smaller than
\begin{align}
E_\omega&=\frac{h}{e v_F t^2} = \frac{h \omega^2}{e v_F}, 
&
B_\omega&=\frac{h}{2e v_F^2 t^2} = \frac{h \omega^2}{2e v_F^2},
\end{align}
where we set the typical time scale to be $t=1/\omega$:
(i) the NLEMRs are non-perturbative with respect to the applied EMFs
when $E$ and $B$ are stronger than these typical scales,
(ii) the NLEMRs can be obtained by a perturbation theory with respect to $E$ and $B$ in the opposite limit.
If we consider the static limit ($\omega \to 0$),
the typical time scale $t=1/\omega$ is replaced by the relaxation time of electrons $\tau$ which is in the order of THz regime, i.e., 
$\tau \simeq 1 \t{ ps}$ for TIs ~\cite{Qu10}.
Magnitudes of electromagnetic fields $E_c$ and $B_c$ 
corresponding to the relaxation time $\tau$ are estimated to be
\begin{align}
E_c& \simeq 80 \t{ V/cm}, 
&
B_c& \simeq 8 \t{ mT}.
\end{align}
Figure \ref{fig: phase diagram} shows the two typical regions of nonlinear responses in the parameter space spanned by $\omega$ and $E$.
NLEMRs show strong nonlinearity in a region bounded by $E>E_\omega$ and $E>E_c$.
NLEMRs show weak nonlinearity in the perturbative region $E<E_\omega$ or in the relaxation-dominant region $E<E_c$.

\section{Strong nonlinearity}
%\textit{Strong nonlinearity ---}
An insight into the strong nonlinearity regime is obtained by considering the static limit $\omega=0$ and $E>E_c$.
In this case, the effective action for static EMFs
is known from $(2+1)$D QED.
The action is given by~\cite{Redlich}
\begin{align}
S&=\int d^3 x L, &
L&=
%\frac{\zeta \left(\frac 3 2 \right)e^{\frac 3 2}}{2^{\frac 5 2}\pi^2} 
A \left(B^2-E^2 \right)^{\frac{3}{4}},
\label{eq: 2+1D QED}
\end{align}
where $A=\zeta \left(\frac 3 2 \right)e^{\frac 3 2}/(2^{\frac 5 2}\pi^2)$
is a numerical constant with $\zeta(x)$ being the zeta function
and we adopted the convention $\hbar=v_F=1$
[Note that $E$ in this unit corresponds to $(c/v_F) E$ in the natural unit ($c=1$)].
This effective action was obtained for the gauge fields with constant $E$ and $B$ fields coupled to a Dirac fermion.
This $(2+1)$D QED action in Eq.~(\ref{eq: 2+1D QED}) is understood from the energy density of Landau levels for Dirac fermions as explained in 
Appendix~\ref{app: 2+1D QED}.
By using this effective action,
the nonlinear responses and current and magnetization 
are summarized as
\begin{align}
J &= i \omega \frac{\partial L}{\partial E}=
- i \frac{3A}{2} \omega E 
\left(B^2-E^2 \right)^{-\frac{1}{4}} 
\\
M &= -\frac{\partial L}{\partial B }=
\frac{3A}{2} B \left(B^2-E^2 \right)^{-\frac{1}{4}},
\label{eq:strong}
\end{align}
which can be translated to the asymptotic 
behavior of the scaling functions $f$ and $g$ as
$f(x,y) \sim x \left(y^2-x^2 \right)^{-\frac{1}{4}}$,
$g(x,y) \sim y \left(y^2-x^2 \right)^{-\frac{1}{4}}$
in the limit of large $|x|$ and $|y|$.
These can be interpreted by the more directly measurable 
quantities, i.e., the dielectric constant $\epsilon$ 
and magnetic susceptibility $\chi$ as
\begin{align}
\epsilon &= -\frac{d^2 L}{d E^2}=
\frac{3A}{4} \left(B^2-E^2 \right)^{-\frac{5}{4}} \left(2B^2-E^2 \right),
\\
\chi &= -\frac{d^2 L}{d B^2}=
-\frac{3A}{4}
\left(B^2-E^2 \right)^{-\frac{5}{4}} \left(B^2-2E^2 \right).
\end{align}
If we focus on the magnetic susceptibility $\chi$,
we can readily see that $\chi$ shows a singularity for $B\to 0$ and $E=0$ as
\begin{align}
\chi &= -\frac{3A}{4} B^{-\frac 1 2},
\end{align}
which indicates that the Taylor expansion with respect to $B$ fields around $B=0$ is not applicable 
and this is a non-perturbative result.
The above formula for $\chi=dM/dB$ satisfies the scaling form for $M$ in Eq.~(\ref{eq: scaling form}) with 
$M \simeq \omega (B/B_\omega)^{1/2}$.
The magnetic susceptibility $\chi$ is subject to a nonlinear suppression with respect to small $E$ as
\begin{align}
\chi &= -\frac{3A}{4} B^{-\frac 1 2} 
\left(1-\frac{3E^2}{4B^2} \right).
\end{align}
When the magnitude of $E$ exceeds that of $B$ ($|E|>|B|$),
the action in Eq.~(\ref{eq: 2+1D QED}) is no longer real 
and indicates that the system becomes unstable.
This corresponds to the Schwinger process where applied electric fields $E$ induce Zener tunneling of an electron between adjacent Landau levels in a successive manner
and the energy distribution of the Dirac electrons is no longer stable.
However, the scaling forms in Eqs.~(\ref{eq: scaling form}), (\ref{eq:strong})
together with Eq.~(\ref{eq: 2+1D QED}) still work. A theoretical study~\cite{Dora} 
on the Schwinger process of the Dirac fermion concluded the
crossover from the linear response regime $J \propto E$ 
to the nonlinear regime $J \propto \tau E^{3/2}$ 
at $E \cong E_c= E_{\omega=1/\tau}$ with $\tau$ 
being the relaxation time. 
Both are consistent with 
Eqs.~(\ref{eq: scaling form}), (\ref{eq:strong}) as
$f(x,0) \sim x$ $(|x| \ll 1)$, $f(x,0) \sim x^{3/2}$  $(|x| \gg 1)$ 
when one replace $\omega$ by $1/\tau$.
In the case of the surface Dirac fermion in TIs, the current $J$ is tied to
the spin polarization $\braket{s}$ due to the spin-momentum locking in TIs.
Thus the Schwinger process of the Dirac fermions in TIs leads to a strong nonlinear spin generation with applied electric fields; we expect the similar crossover for spin generation from 
 the linear response regime  $\braket{s} \simeq E$ 
to the nonlinear response regime  
$\braket{s} \simeq \tau E^{3/2}$ 
with strong $E(\gg E_c)$.

\section{Weak nonlinearity}
%\textit{Weak nonlinearity ---}
When the magnitudes of the applied electromagnetic fields do not exceed the typical scales ($E<E_\omega$, $B<B_\omega$) or they are below the the scales determined by the relaxation time ($E<E_c$, $B<B_c$),
the nonlinear responses show weak nonlinearity 
where perturbative treatments with respect to $E$ and $B$ are valid.

Weak nonlinear responses are described by the expansion 
in terms of the applied $E$ and $B$ fields as
\begin{align}
J=\beta_{n,m}(\omega) B^n(\omega) E^m (\omega),
\end{align}
where we consider Fourier components of $J$, $B$, and $E$ oscillating in frequencies of the order of $\omega$.
The behavior of $\beta_{n,m}(\omega)$ can be determined by the 
scaling forms Eq.(\ref{eq: scaling form}) as
\begin{align}
\beta_{n,m}&\propto
\frac{e^{n+m+1}}{\hbar^{n+m}} \frac{v_F^{2n+m-1}}{\omega^{2n+2m-2}}.
\label{eq: scaling beta}
\end{align}
In a similar manner,
the nonlinear response of the orbital magnetization 
to the applied $E$ and $B$ fields is given by
\begin{align}
M=\chi_{n,m}(\omega) B^n E^m,
\end{align}
and
\begin{align}
\chi_{n,m}& %\simeq L \beta_{n,m}
\propto \frac{e^{n+m+1}}{\hbar^{n+m}} \frac{v_F^{2n+m}}{\omega^{2n+2m-1}}.
\label{eq: scaling chi}
\end{align}

\section{Nonlinear response functions}
%\textit{Nonlinear response functions ---}
Now we study nonlinear response functions in the weak nonlinearity regime. 
We obtain the response functions by the diagram approach and confirm that they 
are consistent with 
the scaling form in Eqs.~(\ref{eq: scaling beta}) and (\ref{eq: scaling chi}).
(The details of the calculations are given in Appendices.)
We start with a nonlinear suppression of the orbital magnetic susceptibility with electric fields.
The orbital magnetic susceptibility $\chi(\omega)$ given by
\begin{align}
M&=\chi_{1,0}(\omega)B,
\end{align}
is obtained from the current-current correlation function
\begin{align}
\chi_{1,0}(\omega) &= 
\left. -\frac{1}{2}\frac{\partial^2}{\partial q_y^2} \braket{j_x j_x}(\omega,\Vec q)
\right|_{\Vec q=\Vec 0},
\end{align}
where $q_y$ is the wavenumber in the $y$-direction.
Its real part is written as
\begin{align}
\t{Re}[\chi_{1,0}(\omega)]&=
-\frac{\pi e^2 v_F^2}{32 h}\delta(\omega).
\end{align}
(For details, see Appendix~\ref{app: orbital magnetic susceptibility}.)
The magnetic susceptibility of Dirac fermions is diverging in the dc limit 
where cutoffs are introduced by the relaxation time or the finite temperature~\cite{McClure,Koshino07}.
Now we consider modulation of the orbital magnetic susceptibility by applying electric fields $E(\omega)$
which is captured by the nonlinear response function
\begin{align}
M(\omega')&=\chi_{1,2}(\omega',\omega)B(\omega') E(\omega) E(-\omega).
\end{align}
Here, the magnetic susceptibility is measured in the frequency $\omega'$ in the presence of laser light of the frequency $\omega$,
and $\chi_{1,2}(\omega',\omega)$ quantifies the change of the magnetic susceptibility proportional to the laser intensity.
In the limit of static magnetic fields ($\omega'\to 0$) for a fixed frequency $\omega$ of the incident light,
the nonlinear response function is given by
\begin{align}
\t{Re}[\chi_{1,2}(\omega',\omega)]&=\frac{\pi e^4 v_F^4}{16 h^3 \omega^4} \delta(\omega'),
\end{align}
where the power of $\omega^{-4}$ is 
the one expected from the scaling analysis
(for derivation, see Appendix~\ref{app: orbital magnetic susceptibility}).
Thus the orbital magnetic susceptibility undergoes a nonlinear suppression by applying the electric fields as
\begin{align}
M(\omega')&=
-\frac{\pi e^2 v_F^2}{32 h}\delta(\omega')
\left( 1-\frac{2 e^2 v_F^2}{h^2 \omega^4}|E(\omega)|^2 \right)
B(\omega').
\end{align}
This is a nonlinear cross correlation effect 
between the magnetic properties and the electric fields.
This nonlinear cross correlation effect is enhanced in the low frequency limit, typically in the terahertz regime.

Next we discuss high harmonic generations which are schematically illustrated in Fig.~\ref{fig: WF}.
The $N$-th harmonic generation is a nonlinear process where the electron absorbs $N$ photons of frequency $\omega$ and emits one photon of frequency $N\omega$.
Here we focus on the third harmonic generation
which is described by 
\begin{align}
J(3\omega)=\beta_{0,3}(\omega) E(\omega)^3,
\end{align}
where $J(3\omega)$ is the induced current with frequency $3\omega$,
and $E(\omega)$ is the electric field of incident light.
\footnote{
The high harmonic generation in graphene has been previously studied by using the semiclassical equation \cite{Mikhailov07} and by using the quantum perturbation theory \cite{Mikhailov14}. 
Here, we apply our scaling arguments to the third harmonic generation and examine it by the quantum perturbation theory.
}
The response function $\beta_{0,3}(\omega)$ is related to
a correlation function of three current operators (as detailed in 
Appendix~\ref{app: high harmonic generation})
and is given by
\begin{align}
\beta_{0,3}(\omega)&=\frac{e^4}{\hbar^3}\frac{v_F^2}{384 \omega^4}.
\label{eq: THG}
\end{align}
This response function is usually denoted as $\chi^{(3)}(\omega)$
and takes a value in the order of $10^{-20}~\t{m}^2/\t{V}^2$
for molecular liquids like CS$_2$ in the terahertz regime~\cite{Hoffmann}. 
In the case of Dirac fermions in TIs,
$\beta_{0,3}(\omega)$ in Eq.~(\ref{eq: THG}) amounts to the order of $10^{-12}~\t{m}^2/\t{V}^2$ for $\omega=1 \t{ THz}$,
if thin films of TIs are stacked in a separation of 10 nm.
This is a significant enhancement compared to the conventional molecular liquids 
which is attributed to the gaplessness of Dirac fermions at the surface of TIs, as has been discussed in the case of graphene~\cite{Wright09,Hendry10}.
We note that  a similar order of $\chi^{(3)}$ was theoretically predicted to stacked layers of two-dimensional electron gas~\cite{Citrin01}.

Finally we discuss the nonlinear modulation of refractive index by an incident light.
This effect is described by the response function
\begin{align}
J(\omega')=\tilde \beta_{0,3}(\omega',\omega) E(\omega')E(\omega)E(-\omega).
\label{eq: nonlinear modulation of refractive index}
\end{align}
Here, $\omega'$ is the frequency of transmitting light for which we are interested in the refractive index,
and $\omega$ is the frequency of incident light that modulates the refractive index.
The response function $\tilde \beta_{0,3}(\omega',\omega)$ is a monotonically decreasing function of $\omega'$ for a fixed $\omega$,
and it takes the asymptotic form 
$\tilde \beta_{0,3}(\omega' \to 0,\omega)=(e^4/\hbar^3)(v_F^2/8\omega^4)$
for the nonlinear modulation of the dc refractive index with a laser light of the frequency $\omega$.
(For details, see Appendix~\ref{app: refractive}.)
Again this power law behavior is consistent with the scaling form 
Eq.~(\ref{eq: scaling beta}),
and hence, its significant enhancement is expected in the terahertz region.
Furthermore, due to the spin-momentum locking in Dirac fermions at the surface of TIs,
the induced current indicates spin generation. 
Therefore, 
in the case of TIs,
the nonlinear modulation of the refractive index leads to a nonlinear Edelstein effect~\cite{Edelstein90},
i.e., nonlinear spin generation induced by strong electric fields, as follows.
When the weak electric field is applied to the surface of a TI, 
there appears spin generation $\braket s$ 
proportional to the applied field $E$.
As $E$ is increased, the above nonlinear refractive index in Eq.~(\ref{eq: nonlinear modulation of refractive index}) indicates the nonlinear spin generation
 $\braket s \simeq E^3$.
If $E$ is further increased, the crossover to the strong nonlinearity regime takes place and the spin generation is governed by the Schwinger process with $\braket s \simeq t E^{3/2}$.

\section{Summary}
To summarize, we have studied the nonlinear electromagnetic responses of 
Dirac fermion from the viewpoint of the scaling laws, which provide a unified view
of various nonlinear phenomena in this system. The nonlinearity increases as
the frequency $\omega$ decreases, and the strong limit of the nonlinearity
can be understood by a Lagrangian for constant electric ($E$) and magnetic 
($B$) fields. The crossover strength $E_\omega$ ($B_\omega$)
of $E$ ($B$) are determined by
$E_\omega=\frac{h \omega^2}{e v_F} $
($B_\omega = \frac{h \omega^2}{2e v_F^2}$), and is of the order of
$E_\omega \simeq 80 \t{ V/cm}$ ($B_\omega \simeq 8 \t{ mT}$)
in the terahertz region, which is accessible experimentally.
The nonlinear response of graphene has been experimentally studied 
in the THz regime up to $E=120 \t{ V/cm}$ \cite{Mics15} 
where nonlinearity of thermalized excited electrons has been observed.
Unfortunately, strong high harmonic generation has not been observed in these experiments, which may be attributed to some extrinsic effects such as disorder effects. Thus observation of scaling behavior of nonlinear responses of Dirac fermions is expected in future experiments.
Furthermore, given the recent advances of experiments on TIs \cite{Chang,Checkelsky2},
Dirac fermions in TIs will offer a platform for novel nonlinear electromagnetic responses.

\begin{acknowledgments}
%\textit{Acknowledgment ---}
We thank N. Ogawa and Y. Tokura for fruitful discussions.
 This work
was supported by 
the EPiQS initiative of the Gordon and Betty Moore Foundation
and by JSPS Grant-in-Aid for Scientific Research
(No. 24224009, and No. 26103006) from MEXT, Japan.
\end{acknowledgments}

\appendix

\begin{widetext}

\section{(2+1)D QED action \label{app: 2+1D QED}}
The (2+1)D QED action for constant gauge strength can be obtained 
by considering Landau levels for Weyl fermions.
The Landau level energy and Landau level degeneracy are given by
\begin{align}
E_{N}&=\t{sgn}(N) v_F\sqrt{2 N \hbar eB}, \\
\frac{1}{2\pi \ell^2}&=\frac{eB}{2\pi \hbar}.
\end{align}
Then the energy density is given by
\begin{align}
H&=\sum_{N=-\infinity}^0 \frac{E_{N}}{2\pi \ell^2} 
=(eB)^{\frac 3 2} \frac{v_F}{2\pi} \sqrt{\frac 2 \hbar} \zeta\left(-\frac 1 2\right)
=\frac{v_F \zeta(\frac 3 2)}{2\pi^2 \sqrt{\hbar}} \left(\frac{eB}{2}\right)^{\frac 3 2},
\end{align}
where we used $\zeta(-\frac 1 2)= \zeta(\frac 3 2)/4\pi$ in the last line.
If we replace $B^2 \to (B^2-E^2)$ for the Lorenz invariance and set $e=\hbar=v_F=1$,
we obtain the effective Lagrangian for $(2+1)$D QED.

\section{Orbital magnetic susceptibility \label{app: orbital magnetic susceptibility}}
We study the orbital magnetic susceptibility,
\begin{align}
M&=\chi(\omega)B.
\end{align}
The magnetic susceptibility is obtained from the current-current correlation function as
\begin{align}
\chi &= -\frac{1}{2}\frac{\partial^2}{\partial q_y^2} \braket{j_x j_x},
\\
\braket{j_x j_x}(i\Omega,\Vec q)&=
\frac{1}{(2\pi)^2} \int d^2\Vec k \sum_{i\omega_n}
\t{Tr}[j_x G(i\omega_n+i\Omega, \Vec k+\Vec q) j_x G(i\omega_n, \Vec k)],
\end{align}
where $q_y$ is the wavenumber in the $y$-direction.
This is understood from $M=\nabla \times J$ and $B=\nabla \times A$.
The Green's function and the current operator for the Weyl fermion are given by
\begin{align}
G(i\omega,\Vec k)&=(i\omega-H)^{-1}, \\
j_i&=\frac{\partial H}{\partial k_i},
\end{align}
with
\begin{align}
H&=k_x\sigma_y-k_y\sigma_x,
\end{align}
where we used the convention $\hbar=v_F=1$.
By using the relation $\partial_{q_y} G=-G \partial_{q_y} G^{-1} G= G j_y G$,
this is rewritten as
\begin{align}
\chi(i\Omega)&=
-\frac{1}{(2\pi)^2} \int d^2\Vec k \sum_{i\omega_n}
\t{Tr}[
j_x G(i\omega_n+i\Omega, \Vec k)j_y G(i\omega_n+i\Omega, \Vec k)
j_y G(i\omega_n+i\Omega, \Vec k) j_x G(i\omega_n, \Vec k)
].
\end{align}
The integration over the angle of $\Vec k$ and the summation of Matsubara frequencies give
\begin{align}
\chi(\omega)&=-\frac{1}{2\pi} \int k d k
\frac{32k^4-52k^2 \omega^2-\omega^4}{8k(4k^2-\omega^2)^3},
\end{align}
after the analytic continuation $i\Omega \to \omega$.
The imaginary part is obtained from the residue at $k=\frac \omega 2$ as
\begin{align}
\t{Im}[\chi(\omega)]=-\frac{1}{32\omega}.
\end{align}
By using the Kramers-Kronig relation,
the real part is obtained as
\begin{align}
\t{Re}[\chi(\omega)]&=
\frac{1}{\pi} P \int^\infinity_{-\infinity}d\omega'
\frac{\t{Im}[\chi(\omega')]}{\omega'-\omega}
=-\frac{\pi}{32}\delta(\omega). 
\end{align}

Now we proceed to the nonlinear effect in $\chi$ in the presence of electric fields by focusing on
\begin{align}
M(\omega')&=\chi_{1,2}(\omega',\omega)B(\omega') E(\omega) E(-\omega),
\end{align}
which is a nonlinear modulation of $\chi(\omega')$ proportional to an intensity of incident light $|E(\omega)|^2$.
From the dimension analysis, this modification is expected to take the form
\begin{align}
\chi_{1,2}(\omega',\omega)&\propto \frac{\chi(\omega')}{\omega^4}.
\end{align}
We verify this by an explicit calculation in the following.
The coefficient $\chi_{1,2}(\omega',\omega)$ is obtained by computing four point correlation functions of the current operator $j_x$,
which are illustrated in Fig.~\ref{fig: diagram S1},
and looking at the contributions proportional to $-q_y^2$.
If we again use the relation,  
$\partial_{q_y} G=-G \partial_{q_y} G^{-1} G= G j_y G$,
this is written as
\begin{align}
\chi_{1,2}(\omega',\omega) &=
\t{Re}\left[ \frac{Q(\omega)}{(-i\omega)(i\omega)} \right]
\\
Q(i\Omega)&=
-\frac{1}{(2\pi)^2} \int d^2\Vec k \sum_{i\omega_n}
\t{Tr}[A_1(i\omega_n, i\Omega, i\Omega')
+A_2(i\omega_n, i\Omega, i\Omega')
+A_3(i\omega_n, i\Omega, i\Omega')],
\label{eq: chi12}
\end{align}
where
\begin{align}
A_1(i\omega_n, i\Omega, i\Omega')&=
j_x G(i\omega) j_x G(i\omega+ i\Omega) j_x G(i\omega) j_x G(i\omega+i\Omega') j_y G(i\omega+i\Omega') j_y G(i\omega+i\Omega'),
\\
A_2(i\omega_n, i\Omega, i\Omega')&=
j_x G(i\omega) j_x G(i\omega- i\Omega) j_x G(i\omega) j_x G(i\omega+i\Omega') j_y G(i\omega+i\Omega') j_y G(i\omega+i\Omega'),
\end{align}
and 
\begin{align}
&A_3(i\omega_n, i\Omega, i\Omega')\n
&=
j_x G(i\omega) j_x G(i\omega+ i\Omega) j_x G(i\omega+ i\Omega+ i\Omega') j_y G(i\omega+ i\Omega+ i\Omega') j_y G(i\omega+ i\Omega+ i\Omega') j_x G(i\omega+i\Omega') \n
&\quad 
+
j_x G(i\omega) j_x G(i\omega+ i\Omega) j_x G(i\omega+ i\Omega+ i\Omega') j_x G(i\omega +i\Omega') j_y G(i\omega +i\Omega') j_y G(i\omega+i\Omega') \n
&\quad 
+
j_x G(i\omega) j_x G(i\omega+ i\Omega) j_x G(i\omega+ i\Omega+ i\Omega') j_y G(i\omega +i\Omega'+ i\Omega') j_x G(i\omega +i\Omega') j_y G(i\omega+i\Omega'),
\end{align}
\begin{figure}
\begin{center}
\includegraphics[width=0.7\linewidth]{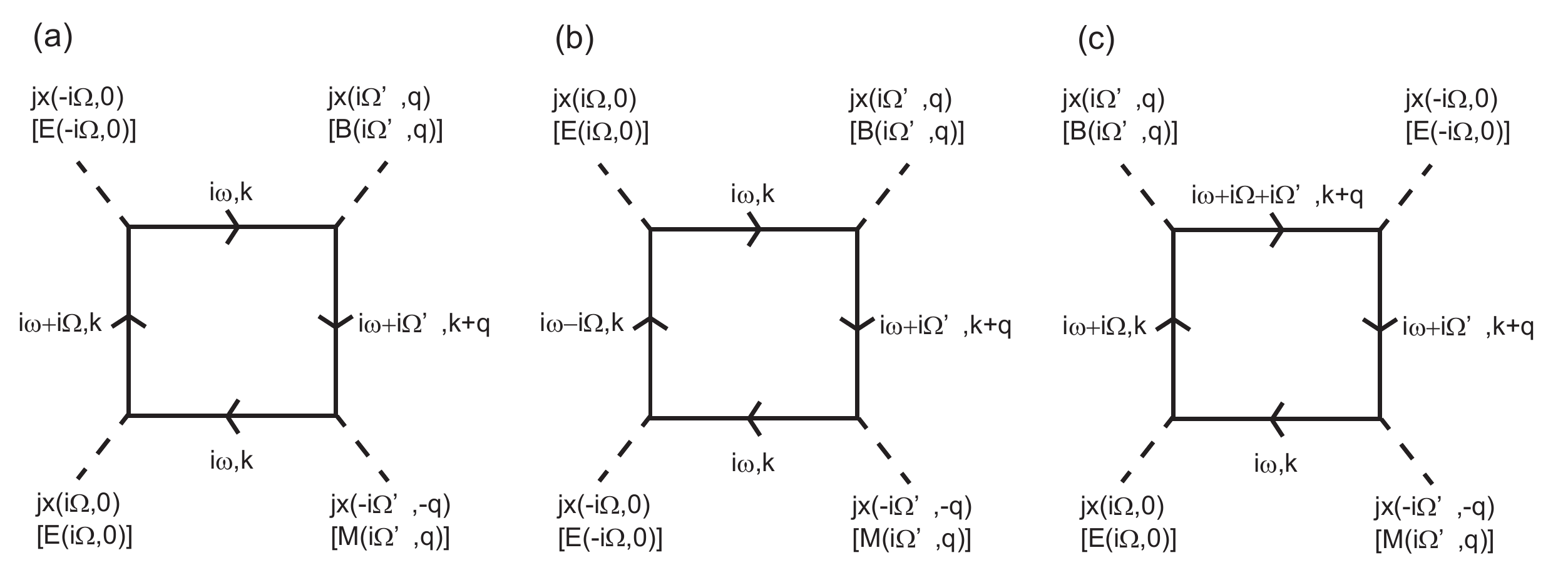}
\caption{\label{fig: diagram S1}
Diagrams for the nonlinear response $\chi_{1,2}(\omega',\omega)$.
Panels (a), (b), and (c) corresponds to $A_1, A_2$, and $A_3$ in
Eq.~(\ref{eq: chi12}), respectively.}
\end{center}
\end{figure}
The result for the imaginary part is given by
\begin{align}
\t{Im}[\chi(\omega)]&=
\begin{cases}
\frac{2\omega^3-2\omega^2\omega'+2\omega\omega'^2+\omega'^3}{32\omega^4 \omega' (\omega+\omega')^3},
& (\omega'<\omega) \\
-\frac{2\omega^3+4\omega^2\omega'-\omega\omega'^2-8\omega'^3}{32\omega \omega'^4 (\omega+\omega')^3}.
& (\omega'\ge \omega) \\
\end{cases}
\end{align}
In the limit of $\omega'\to 0$ for a fixed $\omega$,
this reduces to
\begin{align}
\t{Im}[\chi_{1,2}(\omega',\omega)]&=\frac{1}{16\omega^4\omega'},
\end{align}
which is the form expected from the scaling analysis.
This indicates that the real part of the orbital magnetic susceptibility behaves as
\begin{align}
\t{Re}[\chi(\omega')]&=-\frac{\pi}{32}\delta(\omega')
\left( 1-\frac{2}{\omega^4}|E(\omega)|^2 \right),
\end{align}
which shows a nonlinear suppression of the orbital magnetic susceptibility
with an applied electric field.

\section{High harmonics generation \label{app: high harmonic generation}}
We study the high harmonic generations described by
the nonlinear response,
\begin{align}
J(N\omega)=\chi_N(\omega) E(\omega)^N,
\end{align}
where $J(N\omega)$ is the induced current with frequency $N\omega$,
and $E(\omega)$ is the electric field of incident light.
The coefficient $\chi_N(\omega)$ is given by
an $N$-particle correlation function as
\begin{align}
\chi_N(\omega)&=\t{Re}\left[ \frac{Q(\omega)}{(-i\omega)^N} \right],
\\
Q(i\Omega)&= \frac{1}{(2\pi)^2}\int d^2 \Vec k \sum_{i\omega} 
\t{Tr}\left[j_x G \left(i\omega-i\frac{N-1}{2}\Omega \right) j_x G \left(i\omega-i\frac{N-3}{2}\Omega \right) \ldots j_x G \left(i\omega+i\frac{N-1}{2}\Omega \right)\right]. 
\end{align}

We notice that the coefficients $\chi_N(\omega)$ for even $N$ vanish
after integration over the angle of $\Vec k$
because of the reflection symmetry.
Here we study the third harmonic generation $\chi_3(\omega)$
deduced from
\begin{align}
Q(i\Omega)&=\int d^2 \Vec k \sum_{i\omega} 
\t{Tr} \left[j_x G \left(i\omega-i\frac{3}{2}\Omega \right) j_x G \left(i\omega-i\frac{1}{2}\Omega \right) 
j_x G \left(i\omega+i\frac{1}{2}\Omega \right) j_x G \left(i\omega+i\frac{3}{2}\Omega \right) \right].
\end{align}
After integrating over the direction of $\Vec k$ and summing over the Matsubara frequency,
one obtains
\begin{align}
Q(\omega)&=\frac{1}{2\pi}\int k dk 
\frac{- (k^3 +4 k \omega^2)}
{16 (k^2-\frac 9 4 \omega^2)(k^2-\omega^2)(k^2-\frac 1 4 \omega^2)}.
\end{align}
Since the imaginary part of $Q(\omega)$ is obtained from residues at 
$k=\frac{\omega}{2},\omega,\frac{3\omega}{2}$
(which are evaluated with $\omega$ shifted into the upper half plane),
the coefficient $\chi_3(\omega)$ is given by
\begin{align}
\chi_3(\omega)&=\frac{1}{384 \omega^4},
\end{align}
which is consistent with the scaling law.

\section{Intensity dependent refractive index \label{app: refractive}}
The intensity dependent refractive index is related to
the intensity dependent optical conductivity which is described by
\begin{align}
J(\omega_1)=\chi^{(3)}(\omega_1,\omega_2) E(\omega_1)E(\omega_2)E(-\omega_2).
\end{align}
The coefficient $\chi^{(3)}$ is obtained from a correlation function of four current operators as
\begin{align}
\chi^{(3)}(\omega_1,\omega_2)&=\t{Re}\left[ 
\frac{Q(\omega_1,\omega_2)}{(-i\omega_1)\omega_2^2} 
\right],
\end{align}
and
\begin{align}
Q(i\Omega_1,i\Omega_2)= \frac{1}{(2\pi)^2}\int d^2 \Vec k \sum_{i\omega} 
\t{Tr}\Big[
&
j_x G(i\omega+i\Omega_2) j_x G(i\omega) j_x G(i\omega+i\Omega_1) j_x G(i\omega) \n
&
+j_x G(i\omega+i\Omega_2) j_x G(i\omega) j_x G(i\omega-i\Omega_1) j_x G(i\omega) \n
&
+j_x G(i\omega+i\Omega_2) j_x G(i\omega+i\Omega_1+i\Omega_2) j_x G(i\omega+i\Omega_1) j_x G(i\omega) \Big]. 
\end{align}
After integrating over the direction of $\Vec k$ and summing over Matsubara frequencies,
one obtains
\begin{align}
Q(\omega_1,\omega_2)=\frac{1}{2\pi}\int k dk 
&\frac{-k}{\left(\omega_1^2-4 k^2\right)^2 \left(\omega_2^2-4
   k^2\right)^2 \left(4 k^2-(\omega_1-\omega_2)^2\right)
   \left(4 k^2-(\omega_1+\omega_2)^2\right)}
\n
&
\times
\Big[768 k^8+192 k^6
   \left(\omega_1^2+\omega_2^2\right)-16 k^4 \left(7
   \omega_1^4-29 \omega_1^2 \omega_2^2+7
   \omega_2^4\right) 
\n 
& ~~~~
+4 k^2 \left(\omega_1^6-13
   \omega_1^4 \omega_2^2-13 \omega_1^2
   \omega_2^4+\omega_2^6\right)+5 \omega_1^2
   \omega_2^2
   \left(\omega_1^2-\omega_2^2\right)^2
\Big].
\end{align}
The final result after the $k$ integration reads
\begin{align}
\chi^{(3)}(\omega_1,\omega_2)&=
\begin{cases}
\frac{2\omega_1-\omega_2}{16\omega_1^3 \omega_2 (\omega_1+\omega_2)}, 
& (\omega_1>\omega_2) \\
\frac{2\omega_2-\omega_1}{16\omega_2^4 (\omega_1+\omega_2)}, 
& (\omega_1 \le \omega_2) 
\end{cases}
\end{align}
The coefficient $\chi^{(3)}(\omega_1,\omega_2)$ 
is a monotonically decreasing function of $\omega_1$ for fixed $\omega_2$,
where $\chi^{(3)}\simeq 1/8\omega_1^2 \omega_2$ for $\omega_1 \gg \omega_2$.
In the limit of $\omega_1\to 0$ which corresponds to the dc conductivity modulated by laser light of the frequency $\omega_2$,
one obtains
\begin{align}
\chi^{(3)}(\omega_1 \to 0,\omega_2)&= \frac{1}{8\omega_2^4}.
\end{align}

\end{widetext}

\bibliography{NLEMR}

\end{document}